\documentstyle[preprint,prc,aps,floats,epsf]{revtex}

\tightenlines

\newcommand{\beq}{\begin{equation}}
\newcommand{\eeq}{\end{equation}}
\newcommand{\be}{\begin{eqnarray}}
\newcommand{\ee}{\end{eqnarray}}
\newcommand{\ben}{\begin{eqnarray*}}
\newcommand{\een}{\end{eqnarray*}}

\def\simgt{\stackrel{>}{{}_\sim}}

\newcommand{\wt}{\widetilde}
\def\({\left( }
\def\){\right) }
\def\[{\left[ }
\def\]{\right] }

\newcommand{\Real}{\mbox{Re}\,}

\newcommand{\order}{{\cal O}}

\newcommand{\sing}{$^1\!S_0$ }

\newcommand{\C}[2]{C_{#1}^{(#2)}}
\newcommand{\Lambdac}{\Lambda_c}

\begin{document}
\preprint{\vbox{
\hbox{MIT-CTP-2909}
}}
\vskip1truein

\title{Perturbative Effective Field Theory at Finite Density}

\author{R.J. Furnstahl$^1$, James V. Steele$^{1,2}$, and Negussie Tirfessa$^1$}
\address{$^1$Department of Physics \\
         The Ohio State University,\ Columbus, OH\ \ 43210, USA\\ \bigskip 
	$^2$Center for Theoretical Physics\\Massachusetts Institute of
	Technology, Cambridge, MA\ \ 02139, USA}

\date{October, 1999}

\maketitle

\begin{abstract}
An accurate description of nuclear matter starting from free-space
nuclear forces has been an elusive goal.
The complexity of the system makes approximations inevitable,
so the challenge is to find a consistent truncation scheme with controlled
errors. 
Nonperturbative effective field theories could be well suited for the task.
Perturbative matching in a model calculation is used to explore
some of the issues encountered in 
extending effective field theory techniques to many-body calculations.
\end{abstract}


\thispagestyle{empty}

\newpage

\section{Introduction}

Nuclear forces have been studied in depth over the past fifty years,
leading to excellent phenomenological descriptions~\cite{machleidt}.
A main goal of this enterprise has been to arrive at the ``best''
two-nucleon 
potential and to use it in  many-body calculations to determine the
properties of nuclear matter and finite nuclei.
Effective field theory (EFT)
analyses~\cite{eft} offer a different perspective:
{\it There is no best two-body potential.\/}
Indeed, the off-shell behavior of a potential or amplitude
is not observable and should not
influence the predictability of the theory if a consistent power-counting
scheme is used~\cite{kswDeuteron}.
In addition, many-body contact terms are an inevitable consequence of using
low-energy degrees of freedom and so {\it must\/} be
included along with two-body interactions to eliminate
off-shell ambiguities~\cite{paulo,jimpanic}. 
Finally, the symmetries of the underlying theory of QCD can be
used to constrain the interactions, such as using chiral
symmetry to describe two-pion exchange~\cite{lin,twopi}.
These features 
merit a reanalysis of nuclear matter within the
context of an EFT.

Effective field theory 
techniques may also offer special advantages for many-body
systems, where exact solutions are not realizable.
Estimates of the errors in approximate calculations and clear guidelines for
making systematic improvements are hallmarks of an EFT.
Even with a well-behaved underlying theory, like QED, an EFT can
be beneficial for organizing the contributions from various
effects,
as NRQED does for the radiative corrections to positronium~\cite{nrqed}.
Similar benefits may result from an application of EFT to
nuclear matter.

Skyrme wrote down the first effective theory for nuclear
matter~\cite{skyrme} and used properties of nuclei to fit 
parameters in a mean-field lagrangian.
Early attempts to motivate these parameters from the free-space
nucleon-nucleon potential were promising~\cite{negele}, but
were shown to depend on the form of the two-body potential~\cite{tabakin}.
More recent studies have focused on scaling properties near the Fermi
surface~\cite{shankar,oregon,lutz} to determine which operators are
important for Fermi liquids~\cite{rho} as well as to describe the pairing
gap~\cite{bertsch}.
The behavior of nuclear binding in the chiral limit~\cite{bulgac}
and the nature of the bound state in nuclear matter~\cite{krippa} 
were also examined recently, but without a connection to
free-space parameters.
Here we focus on the use of an effective lagrangian with parameters fit
from free-space and few-body data.

The effective lagrangian consists of long-range interactions constrained by
chiral symmetry and the most general short-range interactions
consistent with QCD symmetries.
The coefficients of these short-range terms may eventually be derived
from QCD, 
but at present must be fit by matching calculated and experimental
observables in a momentum expansion.
In free space, this expansion is in the external momentum over some
breakdown scale $\Lambda$.

For NN scattering, an EFT that consists only of short-range interactions
is equivalent to the effective range expansion, 
which breaks down around $\Lambda \approx m_\pi/2$.
Using chiral symmetry to explicitly include pions in the lagrangian
pushes this scale up to at least $300$~MeV~\cite{lepage,steelef,ksw}, 
but it may  be as high as the $\rho(770)$ mass~\cite{mehen,meissner,steelef2}.
This effective lagrangian can then be used to systematically 
predict other observables, including inelastic processes, up to the
same scale.  
In the following, we discuss extending this procedure to nuclear matter.

If the in-medium expansion parameter
is simply the Fermi momentum $k_F$ over
the free-space breakdown scale $\Lambda$, then $\Lambda \approx 300$~MeV
implies that a useful
EFT description of equilibrium nuclear
matter (where $k_F\approx 260$~MeV) might be impossible.%
\footnote{However, in practice 
numerical factors could improve (or degrade) the situation.}
The analysis of phenomenologically successful mean-field descriptions
of nuclei from an EFT perspective reveals a systematic density expansion, 
with a characteristic mass scale around $600$~MeV~\cite{vmd,meanfield}.
This suggests that a useful EFT expansion of nuclear matter is plausible if
the free-space breakdown scale can be pushed as large as $m_\rho$, but
a determination of how this expansion
carries over to many-body systems is also required.

Important questions about how to
do power counting in a fully nonperturbative many-body
calculation are as yet unresolved.
Nevertheless, we can try to isolate some of the issues and techniques from
free-space EFT analyses to explore how they translate to a many-body system.
Our strategy here is to extend to the medium the {\it perturbative\/} matching
example used by Lepage to illustrate EFT renormalization for 
ordinary two-body scattering~\cite{lepage}.  
This enables us to explore the convergence of non-scattering observables,
the effective in-medium
expansion parameter and breakdown scale, the use of error
plots, and regularization dependence.
Special considerations required due to the large scattering lengths
characteristic of nuclear scattering are postponed to future studies.

The convergence properties of an EFT in the medium can be cleanly tested by
fitting to pseudo-data generated from
an exact underlying potential that features
 a characteristic mass scale.
A full EFT calculation must include long-range pion effects, but these
do not
modify the basic analysis presented here, so
the role of the pion is ignored in order to keep the
discussion as transparent as possible.
Similarly, we omit higher partial waves, spin dependence, 
and relativistic corrections.
For our discussion, 
the only key feature at finite density is the Pauli-blocking of
occupied intermediate states.
With a simple interaction between nucleons that is perturbative in
the coupling, 
expansion parameters in free space and in the medium
can be identified both analytically and using
the corresponding error plots.  

In Sect.~II we perform perturbative matching calculations in
free space using a simple model potential, 
with both dimensional regularization and a cutoff
regularization.
We illustrate the matching procedure in some detail, as it is likely
to be unfamiliar to practitioners of conventional nuclear physics.
In Sect.~III we apply the EFT to finite density and show how
the errors and breakdown scale extend to the medium.
The implications for EFT calculations of nuclear matter
are discussed in Sect.~IV.


\section{Matching in Free Space}

Galilean
invariance requires the interaction between two free nucleons of
momentum ${\bf k_1}$ and ${\bf k_2}$ to be independent of their
center-of-mass momentum ${\bf P}={\bf k_1}+{\bf k_2}$.
For clarity,
we take the underlying potential to be a separable potential in the
\sing-channel%
\footnote{An extension to higher partial waves is
straightforward~\cite{matt} and
does not introduce new features into the analysis.}
that falls off at large momenta and  
depends on the relative momentum
$k \equiv |{\bf k}| \equiv \frac12 |{\bf k_1}-{\bf k_2}|$ only, 
\beq
	\langle {\bf k'}|\hat V_{\rm true}|{\bf k}\rangle = 
		\frac{4\pi}{M}  \frac{\alpha m^3}{(k^2+m^2)(k'{}^2+m^2)} \ .
	\label{true}
\eeq
The mass $m$ corresponds to the range (and non-locality) of the potential;
it plays the role of the underlying short-distance scale.
We adopt an overall normalization of $4\pi/M$, with $M$ the nucleon mass.
The dimensionless coupling $\alpha$ provides a perturbative
expansion parameter.
If $\alpha$ is $\order(1)$, then the effective range parameters
are of ``natural'' size, which means a constant of order one times,
in this case,
a power of $m$ (see Appendix~\ref{appa}).

An EFT can describe observables from the interaction in
Eq.~(\ref{true}) to any desired accuracy 
as long as the details of the underlying potential are not probed.
For scattering observables,
this restricts the external momentum $k$ to be much less than the
characteristic mass $m$.
The effective potential can then be written as a momentum expansion
\beq
  \langle {\bf k'}|\hat V_{\rm EFT}|{\bf k}\rangle
     = C_0 + C_2 \frac{\( k^2+k'{}^2 \)}{2} +
    C_4 \frac{\( k^2 + k'{}^2 \)^2}{4} +
	\wt C_4 \frac{\(k^2-k'{}^2 \)^2}{4} + \cdots \ ,
 \label{veff}
\eeq
which must be regulated.
The two most common regularization schemes are dimensional
regularization with power divergence subtraction (DR/PDS)~\cite{ksw}
and cutoff regularization (CR)~\cite{twopi}.
In the latter case we use 
a gaussian separable cutoff in momentum space that strongly damps
momenta above an arbitrary cutoff $\Lambdac$:
\beq
\! \langle {\bf k'}|\hat V_{\rm EFT}|{\bf k}\rangle = \[ C_0 + 
   C_2 \frac{\( k^2+k'{}^2\)}{2} +  C_4 \frac{\(k^2+k'{}^2\)^2}{4} + 
	\wt C_4 \frac{\(k^2-k'{}^2\)^2}{4} + \cdots \] 
   e^{-(k^2+k'{}^2)/2\Lambdac^2} \ .
  \label{Veffcut}
\eeq
For perturbative calculations, analytic expansions can be obtained for
these schemes in both free-space and in a uniform finite-density system.

The idea of a perturbative matching calculation is that we can work
to $n^{\rm th}$ order in the momentum expansion and determine the
corresponding coefficients $C_0, \ldots, C_{2n}$ order-by-order
perturbatively in the coupling $\alpha$ by equating observables calculated
from the ``true'' 
and EFT hamiltonians \cite{lepage}.
We define the series expansion coefficients $C^{(s)}_{2n}$ by
\beq
C_{2n} = \sum_{s=1}^\infty \alpha^s \,\C{2n}{s} \ .
\label{series}
\eeq
It is not necessary that perturbation theory for the observables converges
in the low-momentum limit,%
\footnote{The main assumption is that the short-distance physics is
perturbative.
The coefficients will have convergent expansions in $\alpha$ since,
as we will see explicitly, higher-order constants incorporate high momentum
physics only and so are not sensitive to possible infrared divergences
(e.g., from a long-range Coulomb potential).  
This means that perturbative matching is sufficient
even if the constants will then be used in a nonperturbative calculation
(see Ref.~\cite{lepage} for an example).}
but for our example it does.
We will carry out the matching for $C_0$ in some detail so
that the generalization to higher orders and the extension to finite density
is clear.
We choose the on-shell $T$-matrix as our matching observable, \hbox{since}
it has a natural perturbative expansion: the Born series
$\hat T = \hat V + \hat V \hat G_0 \hat V + \cdots$, with
$\hat G_0 = (E - \hat H_0)^{-1}$.

\begin{figure}[t]
\begin{center}
\leavevmode
\epsfxsize=5.5in
\epsffile[106 546 498 695]{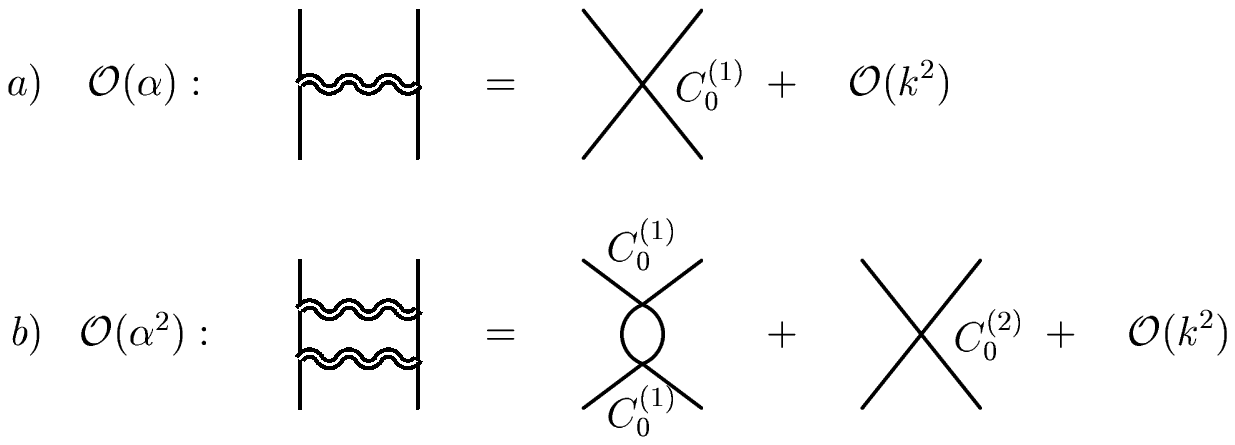}
\end{center}
\caption{Perturbative matching  in free-space to $\order(k^2)$.
The double line for the potential is a reminder that the potential 
is separable.}
\label{fig:diag1}
\end{figure}

We start our discussion with cutoff regularization since the physical
content of the renormalization is manifest.
At $\order(\alpha)$, the on-shell $T$-matrix is just the on-shell
potential $\hat V$, which for  
$k\ll \{m, \Lambdac\}$
can be expanded (it is sufficient to take ${\bf k'} = {\bf k}$ for $S$-waves):
\be
  \langle {\bf k}| \hat V|{\bf k}\rangle &=&  
    \left\{
      \begin{array}{lc}
       \displaystyle
       \frac{4\pi}{M}\frac{\alpha m^3}{(k^2+m^2)^2}
	   = \frac{4\pi\alpha}{Mm}
              + \order(k^2)  , & \qquad\mbox{true} 
       \\ & \\
       \displaystyle
	   \alpha \C01 e^{-k^2/\Lambdac^2} = 
       \alpha \C01 
           + \order(k^2) \ , & \qquad\mbox{EFT (CR)}
        \\
      \end{array}
    \right.
  \label{Vmatch}
\ee
where we have used the shorthand notation $\order(k^2)$ to denote
natural corrections with the proper dimension multiplied by either
$k^2/m^2$ or $k^2/\Lambdac^2$ as applicable. 
Matching to this order fixes
\beq
   \C01 = {4\pi\over Mm} \ ,
   \label{Czeroone} 
\eeq
and is indicated schematically in Fig.~\ref{fig:diag1}a.

Matching at $\order(\alpha^2)$ is illustrated in Fig.~\ref{fig:diag1}b.
There are two contributions on the EFT side:  
$\langle \hat V \hat G_0 \hat V\rangle$ evaluated using $\C01$
fixed by Eq.~(\ref{Czeroone}), and $\langle \hat V \rangle$ evaluated with
the $\order(\alpha^2)$ contribution to $C_0$, namely, $\C02$.
The latter {\it renormalizes\/} $C_0$.
To illustrate the physics of this renormalization, we consider explicitly
the difference of $\langle \hat V \hat G_0 \hat V\rangle_{\rm true}$
and 
$\langle \hat V \hat G_0 \hat V\rangle_{\rm EFT}$:
\be
  \Delta \langle \hat V \hat G_0 \hat V\rangle &=&
  \(\frac{4\pi\alpha }{M m}\)^2
  \Biggl\{
  {m^2 \over k^2+m^2}
  \left[
    \int\! {d^3q\over (2\pi)^3} \, {m^2\over q^2+m^2} {M\over k^2-q^2+i\eta}
	                             {m^2\over q^2+m^2}
  \right]
  {m^2 \over k^2+m^2}   
 \nonumber \\[5pt] 
 & &  \qquad\null -
   e^{-k^2/2\Lambdac^2}
   \left[
    \int\! {d^3q\over (2\pi)^3} \, {e^{-q^2/2\Lambdac^2}} 
	            {M\over k^2-q^2+i\eta}
	            {e^{-q^2/2\Lambdac^2}}
   \right]
   e^{-k^2/2\Lambdac^2} 
   \Biggr\}
  \ . 
   \label{VGVmatch}
\ee
Since we are only working to $\order(k^2)$ in the effective potential, we
can expand all $k^2$ dependence except for $\hat G_0$,
since $q$ can be smaller than $k$:
\be
  \Delta \langle \hat V \hat G_0 \hat V\rangle &=&
  \(\frac{4\pi\alpha }{M m}\)^2
	\int_0^\infty \frac{dq\, q^2}{2\pi^2}
	\, {M \over k^2-q^2+i\eta}
	    \left[
	       {m^4\over (q^2+m^2)^2} - {e^{-q^2/\Lambdac^2}} 
		\right]
    + \order(k^2) 
  \nonumber \\[6pt]
  &=&
-{4\pi \alpha^2 \over M m}
       \left(
          \frac{1}{2} - \frac{1}{\sqrt{\pi}}\frac{\Lambdac}{m} 
       \right)    + \order(k^2) \ .
   \label{VGVmatchb}
\ee
The constant $\C02$ is chosen to cancel the $k^2$--independent part of
$\Delta \langle \hat V \hat G_0 \hat V\rangle$, so that the net
result for $\langle \hat V + \hat V\hat G_0\hat V\rangle$ is $\order(k^2)$:
\beq
   \C02 = -{4\pi \over M m}
       \left(
          \frac{1}{2} - \frac{1}{\sqrt{\pi}}\frac{\Lambdac}{m} 
       \right) \ .
   \label{Czerotwo} 
\eeq
Choosing the cutoff $\Lambdac\approx m$ keeps the constants natural and
gives the maximum range of validity for the EFT~\cite{steelef}.

As expected from the uncertainty principle, the local interaction $\C02$ is 
determined entirely by high-momentum 
($q\simgt\Lambdac$) contributions in the loop integral~\cite{lepage}.
This is because for $q\ll \Lambdac$, the true and EFT tree-level results
have already been matched in Eq.~(\ref{Vmatch}) 
to leading order in $q^2$ and therefore cancel in the
integrand of Eq.~(\ref{VGVmatchb}).
Note  that $\C02$ removes the $\Lambdac$ dependence while correcting
for the contributions in the loop integral from the high-energy states.
The ability to absorb the high-momentum components of the
interaction into the constants of the effective potential
is a generic feature of an EFT
approach and is essential for a systematic prediction.

If we add a long-range potential to the true theory, we must also
reproduce its long-wavelength effects 
in the EFT, so we will still have agreement in the low-momentum
region where $q\ll \Lambdac$, and the analysis goes through unchanged.
If we extend the effective potential to include two constants, 
the $\alpha \C21 q^2$ piece  serves to make the
low-momentum part of the loop integrals agree to $\order(q^4)$,
and the high-momentum part of $\Delta\langle\hat V\hat G_0\hat V\rangle$ 
is absorbed into 
$\alpha^2(\C02 + \C22 k^2)$.
As a result, the EFT reproduces the observables to $\order(k^4)$.
The addition of $C_2$ requires an additional renormalization of $C_0$
in cutoff regularization (denoted by $\delta\C02$, see Appendix~\ref{appb}),
which means that
all the constants change with each successive order in the momentum expansion.

Having convinced ourselves of the physical origin of the constants, we
move to the simpler but physically less transparent dimensional
regularization. 
It has been shown that power counting with only short-range potentials is
regularization scheme independent for free-space 
observables~\cite{steelef,birashort}, as considered here.
Comparing the expressions for the true and effective $T$-matrix to
leading order in $\alpha$ again leads to
\be
    \langle {\bf k}| \hat V|{\bf k}\rangle &=&
    \left\{
    \begin{array}{lc}
        \displaystyle
        \frac{4\pi}{M}\frac{\alpha m^3}{(k^2+m^2)^2}
        = {4\pi\alpha\over Mm} + \order(k^2) 
                 \, , & \qquad\mbox{true} 
        \\ & \\
        \displaystyle
        \alpha \C01 + \order(k^2)\, , & \qquad\mbox{EFT (PDS)}
        \\
    \end{array}
    \right.
\ee
which fixes the first term in Eq.~(\ref{series})
to be
\beq
    \C{0}{1}=\frac{4\pi}{Mm}\ ,
    \label{freec01}
\eeq
just as with the cutoff regulator.

At the next order in $\alpha$, the difference between 
$\langle \hat V \hat G_0 \hat V\rangle_{\rm true}$
and 
$\langle \hat V \hat G_0 \hat V\rangle_{\rm EFT}$ becomes
\be
  \Delta \langle \hat V \hat G_0 \hat V\rangle &=&
  \(\frac{4\pi\alpha }{M m}\)^2
  \Biggl\{
  {m^2 \over k^2+m^2}
  \left[
    \int\! {d^3q\over (2\pi)^3} \, {m^2\over q^2+m^2} {M\over k^2-q^2+i\eta}
	                             {m^2\over q^2+m^2}
  \right]
  {m^2 \over k^2+m^2}   
 \nonumber \\[5pt] 
 & &  \qquad\qquad\qquad\null -
   \left[ \( \frac{\mu}2 \)^{4-D}
    \int\! {d^{D-1}q\over (2\pi)^{D-1}} \, 
	            {M\over k^2-q^2+i\eta}
   \right]
   \Biggr\}
  \ , 
   \label{VGVDRmatch}
\ee
where the second integral is dimensionally regularized with
$D=4+\epsilon$ and in general is evaluated in the PDS scheme as~\cite{ksw} 
\be
\( \frac{\mu}2 \)^{4-D} \int\! \frac{d^{D-1}q}{(2\pi)^{D-1}} \frac{M
k^{2i} q^{2j}}{k^2-q^2+i\eta} = -\frac{M k^{2(i+j)}}{4\pi} (\mu+ik) \ .
\ee
Here $\mu$ is the DR renormalization scale.
As before, the low-momentum parts of the integrands already agree to
$\order(q^2)$, so the constant difference is from the high-momentum behavior.

The results for the first two terms in the
Born series are
\be
    \langle {\bf k}|\hat V + \hat V \hat G_0 \hat V|{\bf k}\rangle
        &=&
        \left\{
        \begin{array}{lc}
            \displaystyle
            \frac{4\pi}{M} \[
              \frac{\alpha m^3}{(k^2+m^2)^2} 
               + \frac{\alpha^2 m^6}{(k^2+m^2)^4}
            \;  \(\frac{k^2-m^2}{2m}-ik\)
            \]   \ ,  
            & \quad\mbox{true}
            \\[8pt]
            \displaystyle
            \alpha \C01 + \alpha^2 \C02 
			               - \frac{\alpha^2 M
		                     \bigl(\C01\bigr)^2}{4\pi} \, (\mu+ik) 
                 \ , 
            & \quad\mbox{EFT (PDS)}
            \\
        \end{array}
        \right.
		\nonumber \\
\ee
which requires 
\beq
    \C{0}{2} = -\frac{Mm}{4\pi} \( \C{0}{1} \)^2 
          \(   \frac{1}{2}  - \frac{\mu}{m}  \) 
         =  -\(\frac{4\pi}{Mm}\) \( \frac{1}{2} - \frac{\mu}{m} \)
           \ ,
    \label{freec02}
\eeq
for a proper match to $\order(\alpha^2,k^2)$.
Note that for $\Lambdac=\sqrt{\pi}\mu$, the DR/PDS constants are
equivalent to the cutoff results Eqs.~(\ref{Czeroone}) and (\ref{Czerotwo}). 
This agreement does not persist at higher orders in momentum.

Carrying the matching to one more order in $\alpha$ suggests a geometric
series for the $\C{0}{s}$ in the DR/PDS scheme, which indeed sums to the full
nonperturbative solution (see Appendix~\ref{appa}):
\beq
    C_0 =  \frac{4\pi \alpha}{Mm}
    \left[ 1 + \alpha\left(\frac{1}{2} - \frac{\mu}{m}\right) \right]^{-1}
 \ .
    \label{c0full}
\eeq
Extending the analysis to higher orders in momentum (or expanding the
full results from Appendix~\ref{appa} in powers of $\alpha$) gives
\beq
    \C{2}{1} = -\(\frac{4\pi}{M m}\)\frac{2}{m^2} \ ,
        \qquad
    \C{2}{2} 
    = 
      \(\frac{4\pi}{M m}\)  \( \frac{5}{8} - \frac{\mu}{m} \)
        \frac{4}{m^2}
      \ ,
    \label{freec2b}        
\eeq
\beq    
    \C{4}{1} = \(\frac{4\pi}{M m}\) \frac{3}{m^4} \ ,
	\qquad
    \C{4}{2} 
       =
       -\( \frac{4\pi}{M m} \)  \(\frac{7}{10} - \frac{\mu}{m}\)
                \frac{10}{m^4}
       \ .   
    \label{freec4b}
\eeq
As additional constants are added to the effective potential in DR/PDS, the 
previously fixed constants are not modified.
This is not the case in CR (see Appendix~\ref{appb}) due to
terms proportional to positive powers of the cutoff.

The constant $\wt C_4$ requires additional input, since it does
not contribute to the on-shell two-body $T$-matrix.
It is tempting to match the true and EFT
$T$-matrices {\it off shell\/} to determine
$\wt C_4$, but this is never necessary, as only observables
are required to fix the EFT constants.
In particular, the additional {\it on-shell\/} constraint of matching
the three-body scattering amplitudes~\cite{paulo} can be used
to find $\wt C_4$.  
Since we are in a regime where the coupling is perturbative, the
Faddeev equations are simplified to a set of three-to-three tree-level
amplitudes at second-order in the potential, $\order(\alpha^2)$.
The result from matching is:  
\beq
 \wt \C{4}{1} = \( \frac{4\pi}{M m} \)  \frac1{m^4}\ .
\label{freec4t}
\eeq
The second-order constant $\wt\C42$ as well as true three-body
terms (needed to absorb the divergences in three-to-three loop-level
amplitudes~\cite{jimpanic})
do not enter nuclear matter calculations until higher
order in the $\alpha$ expansion.

By construction, the EFT systematically reduces the error
order-by-order in the momentum expansion.
We can see this explicitly by examining the truncation error in 
$\langle \hat T\rangle$ to $\order(\alpha^2)$, which is just
$\langle\hat V + \hat V\hat G_0\hat V\rangle$ ,
between calculations using the true
and EFT potentials.%
\footnote{For the present perturbative case, this is more convenient
 than looking at the error in $k\cot\delta$, which is appropriate
 for a nonperturbative calculation.  In the error plots, we
 consider only the real part of the difference in 
 $\langle\hat V + \hat V\hat G_0\hat V\rangle$.
The imaginary part follows from unitarity, which the EFT reproduces
order-by-order.}
For an effective potential only containing $C_0$,
the error is $\order(k^2)$
\be
    \Delta \langle \hat V + \hat V\hat G_0\hat V\rangle
        &=&
            \frac{4\pi}{Mm}
        \left\{
        \begin{array}{lc}
            \displaystyle
			 \biggl[
             \alpha 
                   \biggl(-2 + \frac{m^2}{\Lambdac^2} \biggr) 
                      +  \alpha^2 \biggl(  
                         \frac{5}{2}
                       -  \frac{m^2}{2\Lambdac^2}
                    -\frac{2m}{\sqrt{\pi}\Lambdac}
                   \biggr) \biggr] 
               \biggl(\frac{k}{m}\biggr)^2 
            & \quad\mbox{CR}
            \\[12pt]
            \displaystyle
             \biggl( -2\alpha + \frac{5}{2}\alpha^2 \biggr)
               \biggl( \frac{k}{m} \biggr)^2 
            & \quad\mbox{DR/PDS}
            \\[2pt]
        \end{array}
        \right. 
		\nonumber \\[8pt]
		& & \qquad\null + \order(k^3) \ ,
		\label{errorone}
\ee
and adding $C_2$ brings the error to $\order(k^4)$
\be
    \Delta \langle \hat V + \hat V\hat G_0\hat V\rangle
        &=&
            \frac{4\pi}{Mm}
        \left\{
        \begin{array}{lc}
            \displaystyle
			 \Biggl\{
             \alpha 
                   \biggl(3 - \frac{2m^2}{\Lambdac^2} 
                            + \frac{m^4}{2\Lambdac^4}\biggr) 
                      +  \alpha^2 \biggl[  
                         -7
                       +  \frac{5m^2}{2\Lambdac^2}
                    - \frac{m^4}{4\Lambdac^4}
                  \nonumber \\[8pt]  
                    \displaystyle \null \qquad
                    + \frac{4\Lambdac}{\sqrt{\pi}m}
                      \biggl(
                    1
                    +\frac{m^2}{\Lambdac^2}
                        -  \frac{5m^4}{12\Lambdac^4}
                    \biggr) 
                   \biggr] \Biggr\} 
               \biggl(\frac{k}{m}\biggr)^4 
            & \quad\mbox{CR}
            \\[12pt]
            \displaystyle
            \biggl(3\alpha - 
			7
			               \alpha^2\biggr)
               \biggl(\frac{k}{m}\biggr)^4 
            & \quad\mbox{DR/PDS}
            \\[2pt]
        \end{array}
        \right.
		\nonumber \\[8pt]
      & & \qquad\null   + \order(k^5) \ .
\label{errortwo}
\ee
Note that the truncation error with CR depends on $\Lambdac$ while
in the DR/PDS scheme the truncation error is independent of $\mu$.
Extending the analysis to nonperturbative calculations and finite density
will in general require numerical solutions, 
and so a connection between the analytical results above
and a graphical error analysis is important.
The error plots introduced by Lepage~\cite{lepage} are useful
in this regard.

\begin{figure}[t]
\begin{center}
\hbox{
 \epsfxsize=3.2in
 \epsffile{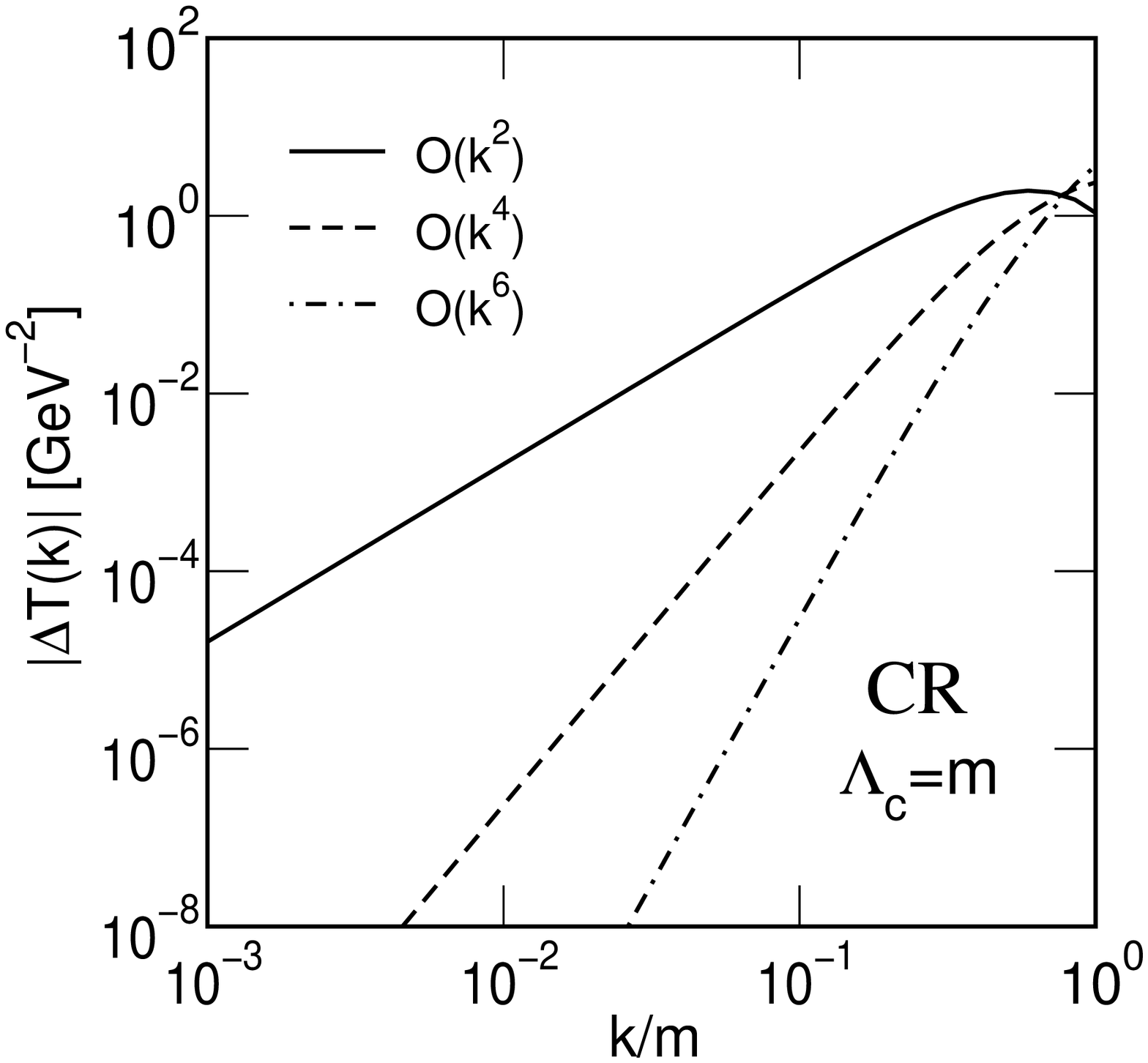}
 \hfill
 \epsfxsize=3.2in
 \epsffile{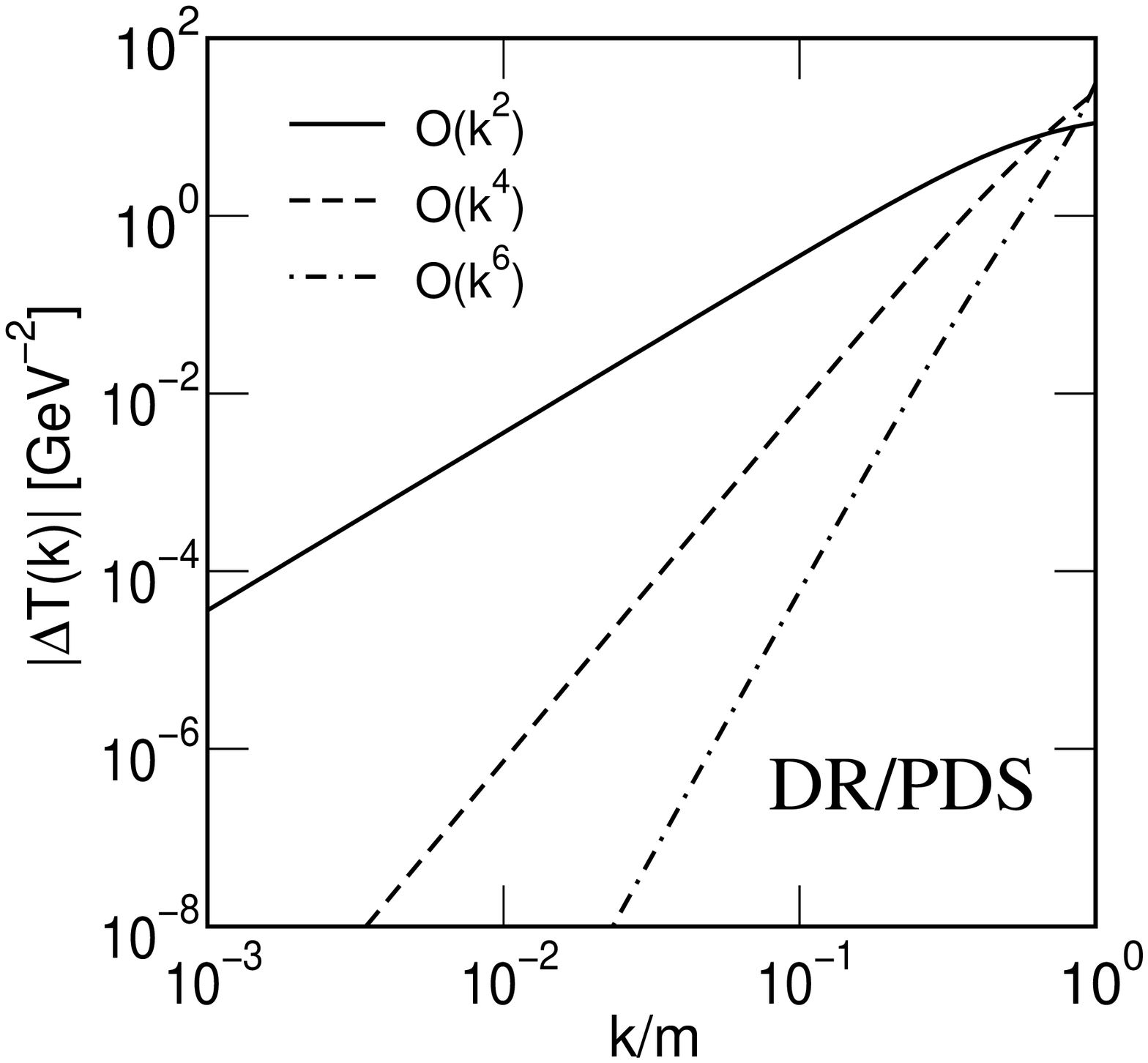}
}
\end{center}
\vspace*{-.5in}
\caption{\label{error}
 $|\Delta\, \Real\!  
 \langle {\bf k}| \hat T | {\bf k}\rangle|$ 
 to $\order(\alpha^2)$ vs.\ $k/m$ for
$\alpha=-1/2$ and with $M=940$~MeV and $m=600$~MeV.  
Results for cutoff regularization (CR) with $\Lambdac=m$
are shown on the left
and for dimensional regularization (DR/PDS) on the right. 
Note that $a_s=-2/3m$
and $r_e=9/m$.}
\end{figure}

Keeping constants to $\order(k^0)$, $\order(k^2)$, and $\order(k^4)$ 
in the effective potential Eq.~(\ref{veff}) or Eq.~(\ref{Veffcut})
leads to successively better approximations to
$\langle\hat V + \hat V\hat G_0\hat V\rangle$, 
as seen by the analytic expressions for
the error Eqs.~(\ref{errorone}) and (\ref{errortwo})
and shown graphically in Fig.~\ref{error}. 
With each additional order, the slope of the error increases,
reflecting the improved truncation error.
(If  a long-range potential is  added to both the true and effective
potentials, the absolute error in the DR/PDS power-counting scheme
will decrease but the error is always $\order(k^2)$~\cite{steelef,KapJim}.)
For $k \ll m$ the error is dominated by the first unmatched term,
so the lines are straight on a log-log plot.
The EFT should break down when the external momentum probes the
details of the underlying potential; graphically 
the intersection of the lines%
\footnote{To determine the intersection scale, one should extend the
lines from the straight regions ($k\ll m$) rather than looking at the
actual intersection.}
indicates the approximate breakdown scale $\Lambda$.
In our example, we see that $\Lambda\approx m$, as expected for a natural
theory [with $\alpha \approx \order(1)$];
indeed, all the errors are of the same order for $k\approx m$.
This breakdown scale does not change as more orders in $\alpha$ are
included in the contact interactions, but the accuracy does
improve.

\section{Finite Density Observables}

Now that the constants are determined to some fixed order in the
momentum and coupling expansion,
the EFT can  be used in a many-body calculation and compared with
 results for the true potential. 
Since the EFT constants are matched to data in free space, 
renormalization-scheme independence requires a comparable level of
sophistication for a calculation in the medium.
If the free-space $T$-matrix is solved using the
nonperturbative Lippmann-Schwinger equation, the in-medium
$T$-matrix (denoted by $\hat\Gamma$) will at least require a 
nonperturbative solution of
the Bethe-Goldstone equation \cite{fw71}. 
However, this leads to many complications at finite density.
For example, the two-nucleon Green's function $\hat G$ now depends
on the interaction 
between the nucleons, which itself depends on $\hat \Gamma$, 
leading to a self-consistent equation.
The integrations over nucleon states are much more complicated since 
occupied intermediate states are Pauli-blocked.
Furthermore, additional terms may also be required, 
such as interactions between particles and
holes.
The development of a consistent yet tractable power counting scheme
for finite density is a critical issue for future investigations.

\begin{figure}[t]
\begin{center}
\leavevmode
\epsfxsize=2.5in
\epsffile{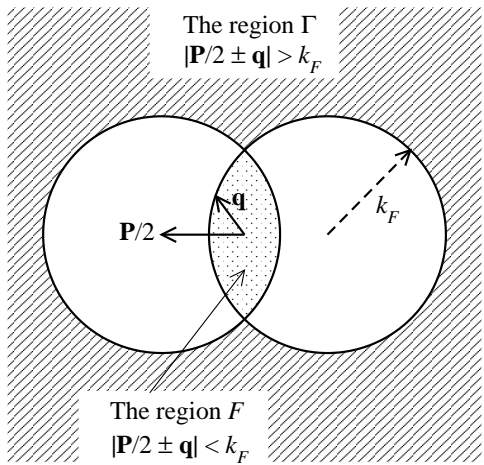}
\end{center}
\caption{Integration regions $\Gamma$ and $F$.} 
\label{fig:fermi}
\end{figure}

Here, however, these issues are greatly simplified by virtue of having 
perturbative interactions and working only to second order in $\alpha$. 
Replacing $\hat G$ with the
free-space propagator $\hat G_0$ 
only changes $\hat \Gamma$ at $\order(\alpha^3)$,
and the perturbative calculation including the Pauli exclusion
projection operator can be performed analytically to this order as
well.
The equation for $\hat \Gamma$ can be written in momentum space as
\beq
    \langle {\bf k'}| \hat \Gamma_{\bf P} | {\bf k} \rangle
    = \langle {\bf k'}|\hat V | {\bf k}\rangle 
    + \int_{\Gamma} \frac{d^3q}{(2\pi)^3} \frac{\langle {\bf k'}| \hat V |
    {\bf q} \rangle \; \langle {\bf q} | \hat V| {\bf k}
    \rangle}
    {(k^2-q^2)/M}
    + \order(\alpha^3) \,,
    \label{bg}
\eeq
with the region of integration $\Gamma$ given by the exclusion of the two Fermi
spheres depicted in Fig.~\ref{fig:fermi}:
\beq
    \int_\Gamma \frac{d^3q}{(2\pi)^3} = \int\! \frac{d^3q}{(2\pi)^3}\,
    \theta\Bigl(\Bigl| {1\over 2}{\bf P}+{\bf q}\Bigr|-k_F \Bigr) \,
    \theta\Bigl(\Bigl| {1\over 2} {\bf P}-{\bf q}\Bigr|-k_F \Bigr) \ .
\eeq
The subscript on $\hat \Gamma_{\bf P}$ in Eq.~(\ref{bg})
is a reminder that, due to the presence
 of the Fermi sea, there is an explicit
dependence on the center-of-mass momentum ${\bf P}$ of the two interacting
nucleons, which to this order
appears only in the integration limits.

The matrix elements of $\hat \Gamma_{\bf P}$ can be used to find the ground
state energy per particle for a system of $A$ nucleons \cite{fw71}:
\beq
    \frac{E_{\rm g.s.}}{A} = 
    \frac35\, \frac{k_F^2}{2M} + \frac{E_{\rm int}}{A}\ ,
     \label{egs}
\eeq
with
\beq
    \frac{E_{\rm int}}{A} = \frac{g}{2\rho} 
    \int_F \frac{d^3P\, d^3k}{(2\pi)^6} 
        \left( g \langle {\bf k}| \hat \Gamma_{\bf P}| {\bf k} \rangle 
        -  \langle {\bf k}| \hat \Gamma_{\bf P}| {\bf -k} \rangle \right) \ .
    \label{eint}
\eeq
Here $g=(2S+1)(2I+1)$ is the spin-isospin degeneracy factor, 
$\rho=g k_F^3/(6 \pi^2)$ is the density of nucleons,
and the integration region
$F$ is the intersection of the two Fermi spheres, as depicted
in Fig.~\ref{fig:fermi}.
We will use $E_{\rm g.s.}/A$ as our basic finite density observable.

We start with the DR/PDS effective field theory.
Evaluating Eq.~(\ref{eint}) to $\order(\alpha)$ is just the Hartree-Fock
approximation;  
the relevant diagrams are shown in Fig.~\ref{fig:diag2}a.
Both the EFT and true results can be analytically integrated 
and expanded (see
Appendix~\ref{appc}) to give (with $\xi \equiv k_F/m$)
\be
    \frac{E_{\rm int}^{(1)}}{A} &=& 
      (g-1) \frac{\alpha m^2 }{3\pi M}
    \nonumber \\[6pt]   & & \null \times
    \left\{
    \begin{array}{lc}
        \displaystyle
        \xi^3 - \frac{3}{5}\xi^5 + \frac{27}{70}\xi^7
          + \order(\xi^{9}) \ , & \quad\mbox{true}
    \\[8pt]
        \displaystyle
        \frac{Mm}{4\pi} \[ \C{0}{1}\xi^3 
           + \frac{3}{10} m^2 \C{2}{1}\xi^5 +
        \frac{9}{70} m^4 \C{4}{1}\xi^{7}\] + \order(\xi^{9}) 
        \ , & \quad\mbox{EFT (DR/PDS)}
\end{array}
\right.
\label{pds1}
\ee
with terms only up to $C_4$ shown in the EFT calculation.
As each new constant $C_{2n}$ is added to the effective potential
and the values found in free space are used
from
Eqs.~(\ref{freec01}), (\ref{freec02}), (\ref{freec2b}), and (\ref{freec4b}),
the EFT result for $E_{\rm g.s.}/A$ 
coincides with the true result up to truncation errors of
$\order(k_F^{2n+5})$ and $\order(\alpha^2)$.
There is no surprise here:
The perturbative matching at finite density must work to this order
since the EFT is
merely parametrizing the Taylor expansion of $V_{\rm true}$ before
the integration over the intersection of the Fermi spheres in Eq.~(\ref{eint}).

\begin{figure}[t]
\begin{center}
\leavevmode
\epsfxsize=6.5in
\epsffile[91 531 555 689]{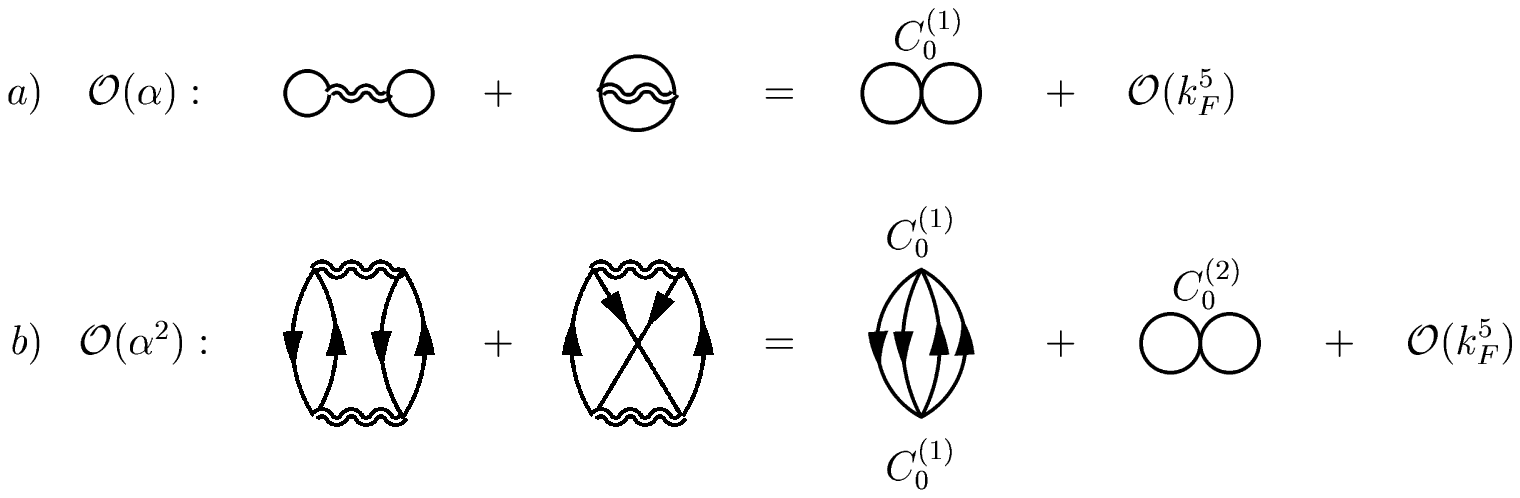}
\end{center}
\caption{Perturbative matching for the energy in nuclear
  matter at $\order(k_F^5)$.}
\label{fig:diag2}
\end{figure}

Taking the analysis to the next order in $\alpha$ 
requires an evaluation of
the Pauli-blocked $\langle\hat V\hat G_0\hat V\rangle$
shown in Fig.~\ref{fig:diag2}b, and hence is a better test of
how the EFT carries over to nuclear matter.
The low-momentum part of the true and EFT
integrals still agree up to $\order(k^{2n})$, 
since the non-Pauli-blocked
part includes the same momentum states in each.
If $k_F \ll m$, the Pauli blocking has no effect on
the high-momentum parts of the integrals, which are therefore the
 {\it same\/} as
in free-space.
Thus the renormalization in free space of $\C{2n}1$ by $\C{2n}2$ 
carries over directly.
Integrating over $F$ to find the energy leaves the same truncation
error, $\order(k_F^{2n+5})$, as at order $\alpha$.

\begin{figure}[t]
\begin{center}
\hbox{
 \epsfxsize=3.2in
 \epsffile{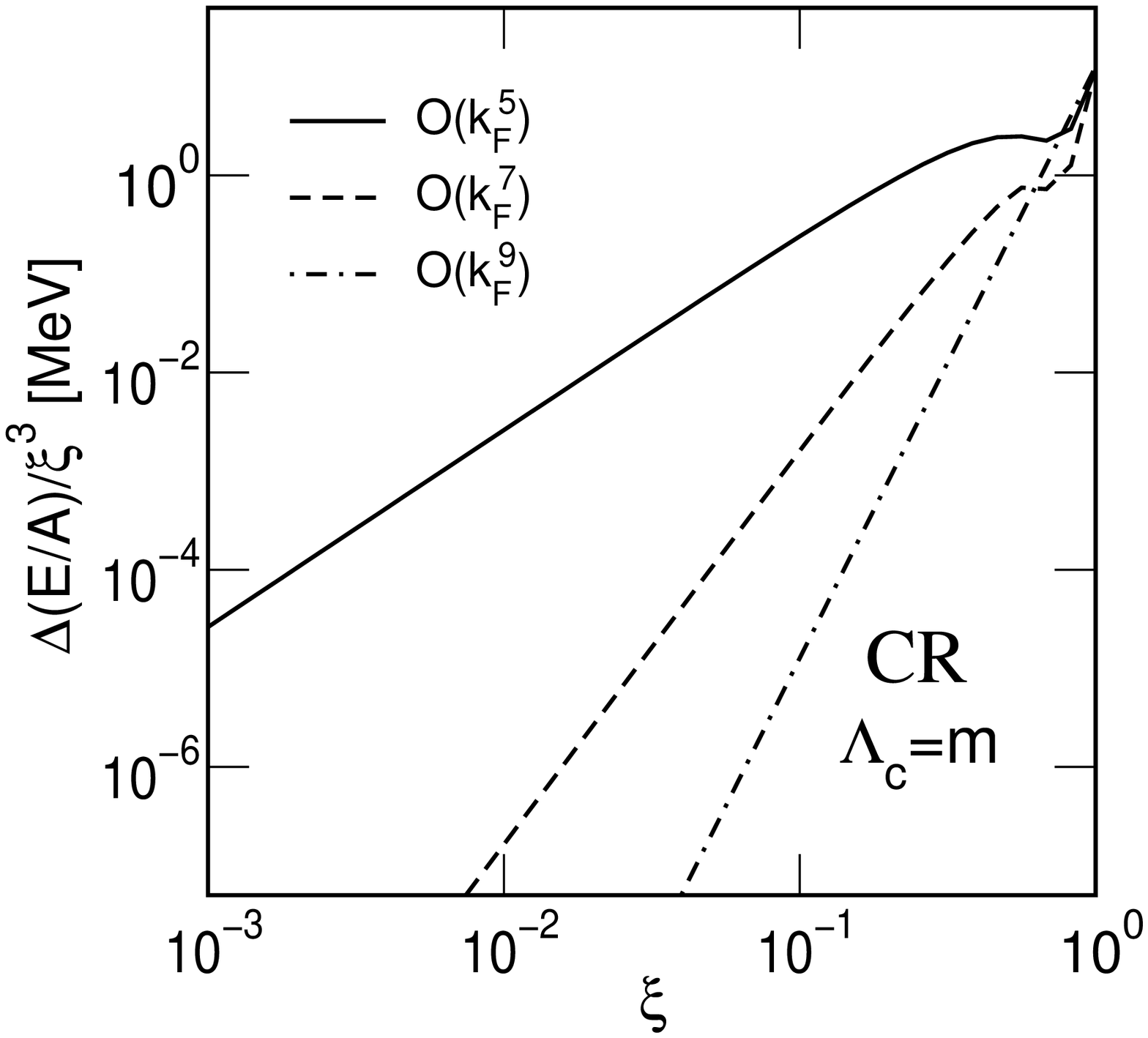}
 \hfill
 \epsfxsize=3.2in
 \epsffile{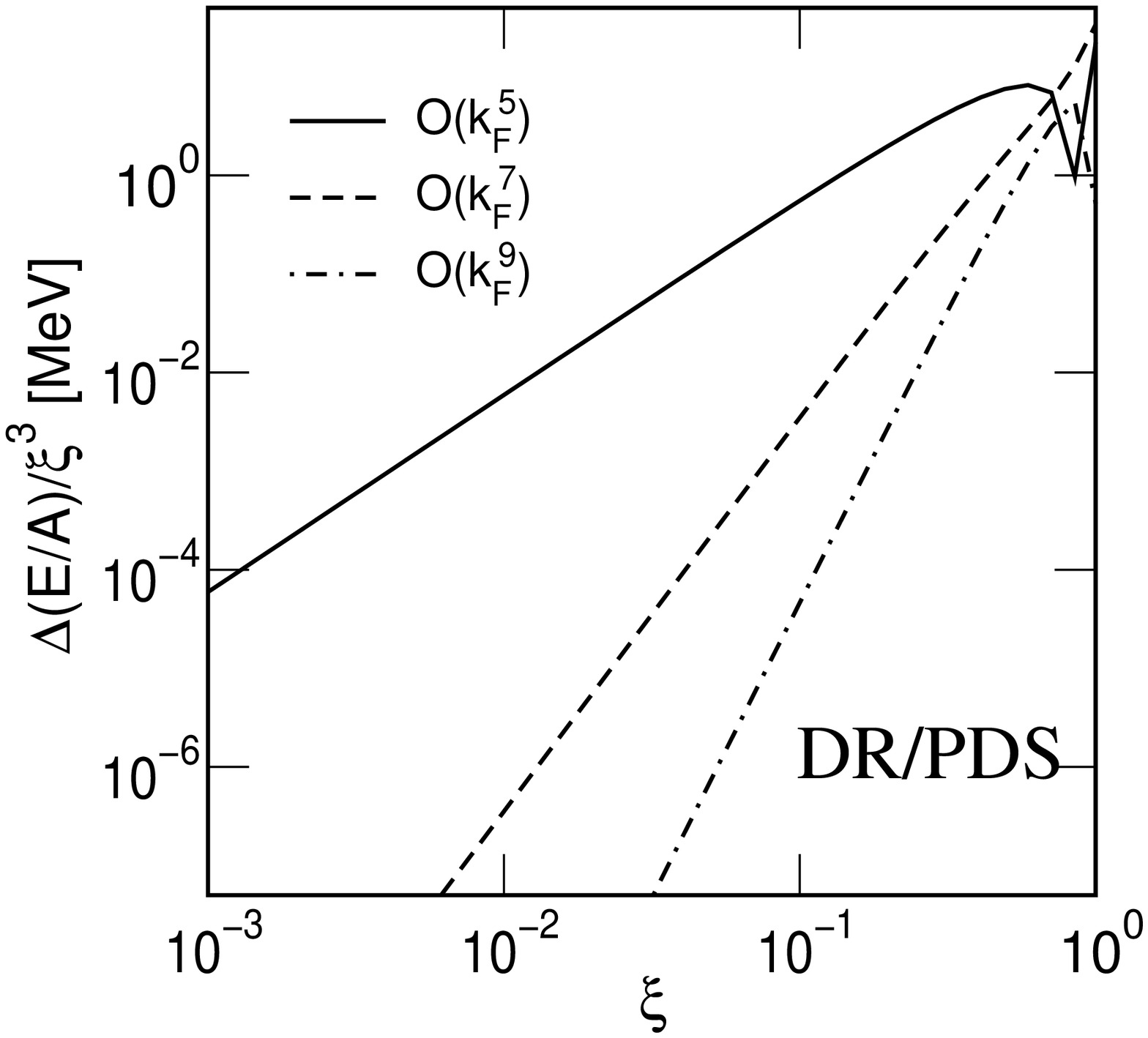}
}
\end{center}
\vspace*{-.5in}
\caption{\label{fig:EoverAerror}
 Error in the ground state energy per particle in nuclear matter
 to $\order(\alpha^2)$ vs.\ $\xi\equiv k_F/m$ for
$\alpha=-1/2$ with $M=940$~MeV and $m=600$~MeV.  
The error is divided by $\xi^3$ for clarity.
Results for cutoff regularization (CR) with $\Lambdac=m$
are shown on the left
and for dimensional regularization (DR/PDS) on the right. }
\end{figure}

The steps involved in actually 
carrying out the integrations are outlined in Appendix~\ref{appc}
and the final expressions are given there. 
If we expand the energy per particle in powers of $\xi$, the true result is 
\be
\mbox{true:}\qquad
    \frac{E_{\rm int}^{(2)}}{A} &=& 
      - (g-1) \frac{\alpha^2 m^2 }{6\pi M}  \label{trueinttwo}
     \\[6pt]   & & \null \times
    \biggl[
        \xi^3 
		- \frac{12(11-2\ln 2)}{35\pi} \xi^4
		- \frac{3}{2}\xi^5 
		+ \frac{16(167/3-8\ln 2)}{105\pi}\xi^6
 \nonumber\\[6pt] & & \quad\null
		+ \frac{9}{5}\xi^7
	- \frac{4(26101/6-470\ln 2)}{1155\pi} \xi^8
          + \order(\xi^{9})
    \biggr]  \ ,
\nonumber
\ee          
and the EFT result using DR/PDS regularization is
\be
\mbox{DR/PDS:}\;\;
    \frac{E_{\rm int}^{(2)}}{A} &=& 
      - (g-1) \frac{\alpha^2 m^2 }{6\pi M}
\label{pdstwo}
     \\[6pt]   & & \null \times
    \biggl\{
        \frac{Mm}{4\pi} \biggl[ 
		    \( -2\C{0}{2} + \frac{M\mu}{2\pi} \(\C01\)^2 \)\xi^3 
		- 3Mm \(\C01\)^2 \frac{(11-2\ln 2)}{35\pi^2} \xi^4 
	 \nonumber \\[8pt]
	  & &  \quad\null
           - \frac35 \( \C22 - \frac{M\mu}{2\pi}\, \C01\C21 \) m^2 \xi^5 
		   - M m^3 \C01\C21 \frac{2(167/3-8\ln 2)}{105\pi^2}\xi^6
     \nonumber
	 \\[8pt]
	   & & \quad\null
        - \frac{9}{35} \( \C{4}{2}
		      - \frac{M\mu}{4\pi} \( \(\C21\)^2 + 2 \C01 \C41 \) 
		  \) m^4 \xi^{7}
	-M m^5 \C01 \wt \C41 \frac{\xi^8}{5\pi^2}
	\nonumber \\[8pt]
	& & \quad\null
	-M m^5 \(2\C01\C41 + \( \C21\)^2\) 
	\frac{(24715/6-470\ln 2)}{11550\pi^2} \xi^8 
		\biggr] + \order(\xi^{9})  \biggr\}
        \ .
\nonumber
\ee
Again, substitution of the expressions up to $C_{2n}$ from
Eqs.~(\ref{freec01}), (\ref{freec02}), and (\ref{freec2b}--\ref{freec4t}) 
removes the $\mu$ dependence and 
leads to agreement with the true result 
up to truncation errors of $\order(k_F^{2n+5})$.
For a perturbative DR/PDS calculation
with $k_F \ll m$, these errors are simply given by the
first unmatched term in the expansion of the true result.
For one constant the leading error is
\beq
  \frac{\Delta E_{\rm int}}{A} = (g-1)\frac{m^2}{\pi M}
       \( -\frac{1}{5}\alpha + \frac{1}{4}\alpha^2 \)
       \xi^5 + \order(\xi^6) \ ,
\eeq
and for two constants it is
\beq
  \frac{\Delta E_{\rm int}}{A} = (g-1)\frac{m^2}{\pi M}
       \frac{3}{10}\( \frac{3}{7}\alpha - \alpha^2 \)
       \xi^7 + \order(\xi^8) \ .
\eeq
The full error is depicted graphically in Fig.~\ref{fig:EoverAerror},
which shows a similar breakdown scale to that in free space.

We emphasize that the contribution from the term in Eq.~(\ref{pdstwo})
proportional to $\wt\C41$ is needed to reproduce the $\xi^8$ term
in Eq.~(\ref{trueinttwo}).
Since this coefficient does not contribute to on-shell two-body 
scattering, it might be interpreted as an ``off-shell ambiguity.''
However, as discussed above, it is determined by on-shell three-body
scattering, so there is no need to consider off-shell effects in
the two-body sector.
At higher order in $\alpha$, we will need full three-body contact terms,
matched in the three-body sector, to account for high-energy
contributions \cite{paulo}. 

We find analogous results with cutoff regularization.
The expanded $\order(\alpha)$ energy is (keeping terms to
$\order(k^2)$ in the effective potential):
\be
\mbox{CR:}\quad    \frac{E_{\rm int}^{(1)}}{A} &=& 
      (g-1) \frac{\alpha m^2 }{3\pi M}
	\left\{  \frac{Mm}{4\pi} \[ \C{0}{1}\xi^3 
           + \frac{3}{10} m^2 \( \C21
		- \frac{1}{\Lambdac^2}\C01 \)\xi^5 
       \] + \order(\xi^{7}) \right\}
\label{eppone}
\ee
and using the explicit expression for $\delta \C02$ from
Appendix~\ref{appb}, the $\order(\alpha^2)$ result is
\be
\mbox{CR:}\;\;    \frac{E_{\rm int}^{(2)}}{A} &=& 
      - (g-1) \frac{\alpha^2 m^2 }{6\pi M} \nonumber \\[6pt] & & \null\times
	\Biggl\{\frac{M m}{4 \pi}
        \Biggl[ \( -2 \C02 +
    \frac{M}{2\pi}\frac{\Lambdac}{\sqrt{\pi}} \(\C01\)^2 \) \xi^3
	- 3Mm \(\C01\)^2 \frac{\(11-2\ln 2\)}{35\pi^2} \xi^4
	\nonumber \\[8pt]
	& & \quad\null
	- \frac35 \Biggl(  \C22 - \frac{\C02}{\Lambdac^2} 
	-\frac{M}{2\pi} 
	\frac{\Lambdac}{\sqrt{\pi}} \[ -\frac3{2\Lambdac^2} \( \C01\)^2 +
    \C01 \C21 + \frac{3\Lambdac^2}{16} \(\C21\)^2 \]\Biggr) m^2    \xi^5
	\nonumber \\[8pt]
	& & \quad\null
	-M m^3 \( \C01 \C21 - \frac{1}{\Lambdac^2}\( \C01 \)^2 \)
    \frac{2(167/3 - 8\ln 2)}{105\pi^2} \xi^6 \Biggr]
		 + \order(\xi^{7})  \Biggr\}
        \ .
\label{epptwo}
\ee
These expressions parallel those found in the DR/PDS scheme in
Eqs.~(\ref{pds1}) and (\ref{pdstwo}).
Taking the values for the constants found from matching 
in free-space (see Appendix~\ref{appb}), and substituting them 
into Eqs.~(\ref{eppone}) and (\ref{epptwo}) 
produces  $\Lambdac$--independent results at each order in the momentum
expansion (although the truncation error depends on $\Lambdac$).
The leading error for one constant is
\beq
  \frac{\Delta E_{\rm int}}{A} = (g-1)\frac{m^2}{\pi M}
       \( 
       \[ -\frac{1}{5} + \frac{1}{10}\frac{m^2}{\Lambdac^2} \] \alpha
       + \[ \frac{1}{4} - \frac{1}{5}\(
              \frac{m}{\sqrt{\pi}\Lambdac} +
              \frac{1}{4}\frac{m^2}{\Lambdac^2} \)
         \] \alpha^2
        \)
       \xi^5 + \order(\xi^6) \ ,
\eeq
and for two constants it is
\be
  \frac{\Delta E_{\rm int}}{A} &=& (g-1)\frac{m^2}{\pi M}
       \frac{3}{10} \Biggl( 
       \[ \frac{3}{7} - \frac{2}{7}\frac{m^2}{\Lambdac^2}
             + \frac{1}{14}\frac{m^4}{\Lambdac^4}
       \] \alpha 
       \nonumber \\  & & \null +
       \[ -1 +
         \frac{4}{7\sqrt{\pi}}\frac{m}{\Lambdac}
         + \frac{5}{14}\frac{m^2}{\Lambdac^2}
         - \frac{5}{21\sqrt{\pi}} \frac{m^3}{\Lambdac^3}
         - \frac{1}{28}
             \frac{m^4}{\Lambdac^4}
           +  \frac{4}{7\sqrt{\pi}}\frac{\Lambdac}{m}
       \] \alpha^2
       \Biggr)\xi^7 
       \nonumber \\ & & \null + \order(\xi^8) \ .
\ee
Just as in the DR/PDS case, the error decreases by a power of
$\xi^{2}$ for each additional order, as shown in Fig.~\ref{fig:EoverAerror}.
This again shows the breakdown scale at finite density is similar to 
the breakdown scale in free space.

In summary, at finite density the energy per particle has similar error
plots to those in free-space when $\Delta E_{\rm g.s.}/A$ 
is plotted against the Fermi momentum $k_F$.
The improvement is by two powers of $k_F$ with each additional
order (although both even and odd powers of $k_F$ are present in the
energy per particle).
The apparent breakdown scale is still $\Lambda\approx m$, without
any  numerical factors significantly different from unity.
The important observation
is not the precise breakdown scale, which depends on the details of the
underlying interaction, but that the breakdown in free
space and in the medium are closely related.
The analysis is easily extended to higher partial waves
for both regularization schemes, which verifies the matching and
the conclusions about the breakdown scale.

\section{Discussion}

The  model we have studied
shows that an EFT expansion for NN scattering 
leads to a low-density expansion of the energy per
nucleon in nuclear matter. 
In particular, in the former one finds scattering observables
order-by-order in the nucleon relative momentum over a breakdown
scale~\cite{eft}.
In the latter, one finds the energy per particle order-by-order
in the Fermi momentum $k_F$ over essentially
the same scale.
We have checked that this basic conclusion is independent of
the details of the model potential.
We expect that this type of expansion will also result when long-distance
pion physics is explicitly included, in analogy to the modified
effective range expansion in free space \cite{steelef}.

Of course, since this is a perturbative calculation it can
only be suggestive.
We will need to clarify a variety of interrelated issues:
nonperturbative power counting in the medium,
the impact of pions, the elimination of off-shell ambiguities, the
role of three-nucleon interactions.
However, if we assume the basic results here carry over
to a full nonperturbative treatment, we can speculate
on how well a leading order (LO) and next-to-leading order (NLO) EFT
 calculation will do in reproducing nuclear matter saturation.

In particular,
we can use phenomenological
energy functionals of nuclear matter (that is,
the energy expressed as a function of the density) to anticipate
what a full EFT calculation might achieve.
Nonrelativistic and relativistic  mean-field
functionals provide parameterizations of the energy per particle
near the equilibrium density, with individual contributions
characterized by a density expansion parameter of about 1/5 \cite{rjfpanic}.
These functionals are fit directly to finite density data.
As such, they include (approximately) the effects of three-body forces and
short-range correlations, even though the formalism is that of Hartree or
Hartree-Fock \cite{rjfpanic}.
(There are certainly important limitations in the descriptions; 
for example, there is no $k_F^4$
term in nonrelativistic mean-field energy functionals.)

If we assume
that the expansions about $k_F=0$ are reasonably determined
by fits near $(k_F)_{\rm equilibrium} < \Lambda$,
then we can use them to predict how far in a truncated $k_F$ expansion
we need to go to see saturation.%
\footnote{A nonperturbative EFT calculation would include contributions
to all orders in $k_F$ and therefore may do even better.}
Ordinary Skyrme Hartree-Fock functionals are truncated at $\rho^3$, which
is $\rho^2 \propto k_F^6$ in the energy per particle.  As more terms
are added to the functional, it should become a more reliable
``extrapolation'' of an expansion about $k_F=0$.
So we can look at a generalized Skyrme model, fit out to $k_F^9$
but then truncated at $k_F^6$, to see how saturation changes.
In this case, the equilibrium density decreases by about 20\%
while the binding energy decreases by roughly 5~MeV \cite{hackworth}.
There is an energy minimum
when only terms up through $k_F^5$ is kept, but it is far removed from the
empirical point.
Thus we expect a qualitative reproduction of saturation if
we keep terms up through $k_F^6$, as long as the effective $\Lambda$
is sufficiently high.

A similar conclusion is reached by examining relativistic mean-field
point-coupling
models.  These models contain contributions to all orders in the
$k_F$ expansion.  One finds that a saturation minimum is first achieved
with an expansion truncated at $k_F^5$ but that the $k_F^6$ term is needed to
get reasonably close. 
The quantitative results again suggest that an EFT calculation up to $k_F^6$ 
might
reproduce nuclear saturation with the equilibrium binding energy
given to within about 5~MeV and the density to within about
20\%.

The mass scale that sets the convergence of contributions to mean-field
energy functionals is of order 600\,MeV \cite{rjfpanic}.
However, this is not necessarily the same as the breakdown scale.
If one  constructs error plots for relativistic mean-field
functionals by comparing the full energy as a function of
$k_F$ to the same functional truncated at $k_F^5$, at $k_F^6$, and so on,
a breakdown scale for the expansion about $k_F=0$ is identified.
For point-coupling models, this breakdown scale is close to 400\,MeV
but for models with meson fields it is actually below 
the equilibrium $k_F$ \cite{rjfbds}.
This discrepancy is not yet understood. 

One might ask how saturation is realized in such a density expansion.
The existence of a minimum in the energy per particle suggests
that different orders in the
expansion must be playing off each other;
a natural conclusion is that this should only occur near the breakdown
scale (where all terms contribute equally) and not at $k_F\ll \Lambda$.
The mean-field phenomenology suggests a resolution:  there {\it is\/} 
an interplay
of different orders but it is highly restricted.  In particular,
repulsion from
the $k_F^5$ and $k_F^6$ terms becomes important compared
to the attraction from the $k_F^3$ piece well below the breakdown
scale $\Lambda$.
Furthermore, this happens because the coefficient of the $k_F^3$ term
is unnaturally small;%
\footnote{Here ``small'' means roughly half the size one might expect;
it is  not fine-tuned to be close to zero.  
If the coefficient is doubled in size, equilibrium does occur
at $k_F \approx \Lambda$ and bound by 200 to 300~MeV.}
in relativistic models this is a direct result of cancellations between
Lorentz scalar and vector contributions that are each of natural size.
The cancellations leading to a small $k_F^3$ term do not recur in higher
orders, so further contributions just correct the equilibrium properties
and do not change the qualitative behavior.

The convergence of the density expansion is also consistent with the
convergence of a conventional expansion of the Galitskii equation in powers
of $k_F$ times the radius of a strong, short-range (hard-core) 
potential\cite{fw71}, which we can associate with $1/\Lambda$.
In the effective field theory treatment, this short-distance expansion 
parameter arises naturally when long-distance pion physics is treated
explicitly.
In free space, consistently including the pion takes us from an effective
range expansion with a breakdown scale of order the pion mass to
a modified effective range expansion with a breakdown scale set by
the next important mass scale.
We expect an analogous situation at finite density.

So what are the ingredients of a minimal EFT
calculation of nuclear matter saturation?
First, we need the breakdown scale for NN scattering in free space to
be as large as possible. 
There are indications that a complete NLO calculation 
using cutoff regularization and including two-pion
irreducible contributions (and possibly the NNLO two-pion pieces) may
achieve $\Lambda\approx 600\,\mbox{MeV}$~\cite{meissner}.
Then we must include all contributions that will generate terms 
in the energy per particle up to 
$k_F^6$.  This implies including 
two constants in each S-wave, one in each P-wave,
and an S--D mixing term. 
One- and two-pion exchange should be included in all waves.
Finally, a three-body contact term will be needed~\cite{paulo}. 
Note that this recipe includes
the standard ingredients of nuclear matter phenomenology:
the pion tensor force, velocity dependence of the interaction,
mid-range attraction from two-pion exchange,%
\footnote{Note that the NLO or NNLO two-pion potential does not include
contributions from two pions interacting ``in flight,'' which is counter to
the usual intuition from nuclear phenomenology about the mid-range
attraction.\/} 
and three-body forces.
While much remains to be done before a consistent nonperturbative
calculation is available, we are optimistic that our speculations will
be tested in the near future.

\acknowledgments

We thank B.~D.\ Serot for useful comments and discussions.
This work was supported in part by the National Science Foundation
under Grants No.\ PHY--9511923 and PHY--9800964.  J.S.\ is also
supported in part by the U.S. Department of Energy under cooperative
research agreement \#DF-FC02-94ER40818.

\appendix

\section{Nonperturbative Free-Space Solutions}\label{appa}

Scattering from the $S$--wave separable potential 
\beq
	\langle {\bf k'}|\hat V_{\rm true}|{\bf k}\rangle = 
	\frac{4\pi m}{M}
      \frac{\alpha m^2 }
		{(k^2+m^2)(k'{}^2+m^2)} \equiv
		V_0(k,k')  \ ,
\eeq
can  be solved analytically to give 
\beq
T_0^{-1}(k,k) = V_0^{-1}(k,k) + \frac{M}{4\pi}\( ik - \frac{k^2-m^2}{2m}\) 
\eeq
or
\beq
	k\cot\delta_0 = -\frac{4\pi}{M V_0(k,k)} + \frac{k^2-m^2}{2m} \ .
\eeq
Expanding $k\cot\delta_0$ in powers of $k^2$ gives the effective
range expansion for this potential:
\beq
   k\cot\delta_0 = -m\( \frac{1}{2} + \frac{1}{\alpha} \)
           + \frac{1}{m} \( \frac{1}{2} - \frac{2}{\alpha} \)k^2
           -  \frac{1}{\alpha m^3} k^4  \ .
\eeq
The effective-range parameters are natural (i.e., characterized by the
underlying mass scale) for $\alpha = \order(1)$.

Regularizing the EFT with DR/PDS gives
\beq
	k\cot\delta_0^{\rm PDS} = - \frac{4\pi}{M V_0^{\rm PDS}(k,k)} -\mu \ ,
\eeq
which can readily be expanded in momentum and solved for the $S$--wave
low-energy
constants $C_{2n}$:
\beq
    C_0 = \frac{4\pi \alpha}{Mm}
    \left[ 1 + \alpha\left(\frac{1}{2} - \frac{\mu}{m}\right)
    \right]^{-1}\!\!\!\!\!\!\ ,
	\;\;\;
	C_2 = - \frac{C_0^{\; 2}}{\alpha}  \frac{M}{2\pi m}   \left( 1 -
		\frac{\alpha}{4}\right)\ ,
	\;\;\;
	C_4 = \frac{C_2^{\; 2}}{C_0} -
		\frac{C_0^{\; 2}}{\alpha}  \frac{M}{4\pi m^3} \ .
\eeq
Expanding these results in powers of $\alpha$ gives the results in 
Eqs.~(\ref{freec01}), (\ref{freec02}), and (\ref{freec2b}--\ref{freec4b}).
Matching the tree-level on-shell three-body scattering amplitude \cite{paulo} 
to leading order in $\alpha$ gives $\wt C_4 = C_0/m^4$.
%


\section{Two Constants with Cutoff Regularization}\label{appb}

For two terms in the effective potential,
the expression for $\C{0}{2}$ [Eq.~(\ref{Czerotwo})] gets an extra piece:
\beq
	\delta \C{0}{2} = \frac{M}{4\pi} \frac{\Lambdac^3}{2\sqrt{\pi}}
	       \left(  \C01\C21 + 
		                {3\over 8}\bigl(\C21\bigr)^2 \Lambdac^2 
            \right) \ ,
\eeq
with
\beq
	\C{2}{1} = -\(\frac{4\pi}{Mm}\) \frac{2}{m^2}
	           + \frac{\C{0}{1}}{\Lambdac^2} \ .
\eeq
The $\order(\alpha^2)$ piece of $C_2$ has more terms compared to the
DR/PDS result Eq.~(\ref{freec2b}):
\be
	\C{2}{2} &=& \(\frac{4\pi}{Mm}\) \frac{5}{2m^2}
	-\frac{M}{4\pi} \frac{\Lambdac}{\sqrt{\pi}}
          \( \frac{3}{\Lambdac^2} \(\C01\)^2 
          -2 \C01\C21 - \frac38 \(\C21\)^2\Lambdac^2 \)
     + \frac{\C{0}{2}}{\Lambdac^2}
		  \ .
\ee
%



\section{Nuclear Matter Integrals}\label{appc}

For an even-parity potential, 
Eqs.~(\ref{bg}) and (\ref{eint}) give
the first-order contribution to the energy per particle as
\beq
	\frac{E_{\rm int}^{(1)}}{A} = \frac{g\,(g-1)}{2\rho} \int_F \frac{d^3P\,
		d^3k}{(2\pi)^6} \, 
		\langle {\bf k}| \hat V|{\bf k}\rangle
	\equiv \frac{g\,(g-1)}{\pi^4 \rho} \, m^4{\cal I} \ . 
\eeq
For an odd-parity potential, the
degeneracy pre-factor  becomes $g(g+1)$.
Introducing scaled momenta ${\bf s}\equiv\frac12 {\bf P}/k_F$
  and ${\bf t}={\bf k}/k_F$, then
\beq
	{\cal I} = \xi^4\int_0^{1} ds\, s^2 \int_0^1 dx 
		\int_0^{z_-(x)} dt\, t^2 \, {\cal F}(t) \ ,
	\label{Ieq}
\eeq
with 
\beq
	z_\pm(x) \equiv \pm sx+\sqrt{1-s^2(1-x^2)}\ ,
		\qquad\qquad
		{\cal F}(t) \equiv k_F^2\,
        \langle k_F{\bf t} |\hat V|k_F{\bf t}\rangle \ .
\eeq
Integrating by parts with respect to $x$ and then $s$,
and changing variables simplifies the integral to
\beq
	{\cal I} = \frac{\xi^4}{6} \int_0^1\!\! ds\, s^2 \[ 
     2 - 3 s + s^3 \] {\cal F}(s) \ ,
	\label{fo}
\eeq
which for the potential in Eq.~(\ref{true}) gives
\beq
	{\cal I} = \alpha
	 \frac{2\pi m}{3M}
	   \[ \xi^2 + \xi^3 \arctan\xi -
		\(1+\frac32 \xi^2\) \ln\(1+\xi^2\)\] \ .
\eeq
The integral can also be directly evaluated for the DR/PDS and
CR effective potentials.

The second-order contribution $E_{\rm int}^{(2)}/A$
follows from the same analysis as above,
but with
\beq
	{\cal F}(t) = k_F^3\; {\cal P}\,\int_\Gamma\! 
		\frac{d^3u}{(2\pi)^3} \langle k_F{\bf t}| \hat V|k_F{\bf u}\rangle 
		\frac{M}{t^2-u^2} \langle {k_F\bf u}| \hat V|k_F{\bf t}\rangle 
        \equiv 
		\int_0^1 dy
		\int_{z_+(y)}^\infty\! du\,
		u^2 {\cal H}(t,u)
		\ ,
	\label{Feq}
\eeq
where
\beq
	{\cal H}(t,u) = \frac{Mk_F^3}{2\pi^2}
	\langle k_F{\bf t}| \hat V|k_F{\bf u}\rangle 
	\frac{1}{t^2-u^2} \langle {k_F\bf u}| \hat V|k_F{\bf t}\rangle \ .
\eeq
We divide the integrals into two parts:
\beq
	{\cal I} =    {\cal I}_1-{\cal I}_2  \ ,
\eeq
with 
\be
	{\cal I}_1 &=& \int_0^1ds\, s^2
		\int_0^1dx\int_0^{z_-(x)}dt\,
		t^2\int_0^\infty du\, u^2 {\cal H}(t,u) \ ,
	\\[6pt]
	{\cal I}_2 &=& 
		\int_0^1ds\, s^2\int_0^1dx\int_0^{ z_-(x)}dt\,
		t^2\int_0^1dy\int_0^{z_+(y)} du\, u^2 {\cal H}(t,u) \ .
\ee
The first integral ${\cal I}_1$ can be evaluated as in Eq.~(\ref{fo}).
For the potential Eq.~(\ref{true}), this leads to
\be
	{\cal I}_1 &=& \alpha^2
	\(\frac{\pi m}{6M}\)
	\[ \ln\(1+\xi^2\) -
		\frac{\xi^2\(2+\xi^2\)}{2\(1+\xi^2\)} \]  
     \nonumber   \\[6pt] 
     &=&  \alpha^2
	\(\frac{\pi m}{6M}\) \[ -\frac{\xi^6}{6} + \frac{\xi^8}{4} 
           - \frac{3\xi^{10}}{10} + \order(\xi^{12})  
        \]
        \ .
\ee
The second integral ${\cal I}_2$ is not conveniently expressed in closed
form.
For the potential Eq.~(\ref{true}), it can be expanded in $\xi$ to give
\be
	{\cal I}_2 &=& \alpha^2 	\(\frac{m}{105M}\)
           \biggl[- \xi^7 (11-2\ln 2) +
		\frac{4\xi^9}{9} \(\frac{167}{3} - 8\ln 2\) 
\nonumber\\
& & \qquad\null
	- \frac{\xi^{11}}{99} \( \frac{26101}{6} - 470 \ln 2\)
	+\order(\xi^{13})\biggr] \ .
\ee
Results for the EFT are found similarly.

\end{document}